# PPG-Based Heart Rate Accuracy in Diverse Populations: Investigating Inequities Across Body Composition and Skin Tones

Kostrna, Jason (corresponding)

Florida International University, Department of Teaching and Learning, 11200 SW 8th Street, Miami, FL 33199, jkostrna@fiu.edu

https://orcid.org/0000-0002-4838-7094

Oparina, Ekaterina

Florida International University, Department of Teaching and Learning, eoparina@fiu.edu

https://orcid.org/0000-0002-0810-4889

Ramella-Roman, Jessica C.

Florida International University, Department of Biomedical Engineering,

jessica.ramella@fiu.edu

https://orcid.org/0000-0002-5710-6004

Palacios, Cristina

Florida International University, Department of Dietetics & Nutrition, cristina.palacios@fiu.edu

https://orcid.org/0000-0001-9437-0376

Rodriguez, Andres J.

Florida International University, Department of Biomedical Engineering, andrrodr@fiu.edu

https://orcid.org/0000-0002-2159-8123

Pei, JunZhu

Florida International University, Department of Biomedical Engineering, jpei@fiu.edu

https://orcid.org/0000-0002-8462-0893

Ajmal, Ajmal

Florida International University, Department of Biomedical Engineering, aajmal@fiu.edu

https://orcid.org/0009-0000-4600-0602

The authors report there are no competing interests to declare.

This research was supported by a pilot grant from the National Science Foundation (NSF) via the Precise Advanced Technologies and Health Systems for Underserved Populations (PATHS-UP) program.

**Abstract**

Wearable devices are widely used for heart rate (HR) monitoring, yet their accuracy across diverse body compositions and skin tones remains uncertain. This study evaluated four wrist-worn devices (Apple, Fitbit, Samsung, Garmin) in 58 Hispanic adults with Fitzpatrick skin types III–V during a cycling protocol alternating moderate (64–76% HRmax) and vigorous (77–95% HRmax) intensities. Criterion HR was obtained via Polar H10 ECG, with accuracy assessed using mean absolute error, mean absolute percentage error (MAPE), bias, and intraclass correlation coefficients. All devices significantly deviated from criterion measures ($p < .001$). Apple and Garmin showed the lowest error, while Fitbit and Samsung produced greater inaccuracies. Higher BMI and darker skin tones increased MAPE. These biases disproportionately affect higher-risk groups, underscoring the need for improved algorithms to ensure equity in health monitoring.

# Introduction

Accurate heart rate (HR) measurement is essential for monitoring cardiovascular health and estimating physiological workload and energy expenditure (Etiwy et al., 2019; Garet et al., 2005; Ludwig et al., 2018). HR is used to prescribe training zones, track fitness, monitor health, and guide clinical recommendations (Povea & Cabrera, 2018). In research and real-world applications, photoplethysmography (PPG) has become the dominant method for capturing HR in wrist-worn wearable devices (Charlton & Marozas, 2022; Kim & Baek, 2023). While PPG offers practical advantages (e.g., compact sensors, non-invasive measurements, and continuous data collection) it remains susceptible to error due to its reliance on optical signals and underlying assumptions about tissue properties (Kim & Baek, 2023). These errors can take multiple forms, including systematic bias (i.e., consistent over- or underestimation), absolute error (i.e., deviation from the criterion measure), proportional inaccuracy (e.g., mean absolute percentage error), intermittent signal dropout, or reduced reliability across repeated readings (i.e., low intraclass correlation; Kim & Baek, 2023). Importantly, these sources of error appear to be more pronounced in certain populations, including individuals with darker skin pigmentation or higher body mass, raising concerns about the accuracy of these tools (Shcherbina et al., 2017). When HR measurement is flawed, the downstream consequences ripple across multiple domains. PPG-derived HR is commonly used to calculate resting heart rate, estimate basal metabolic rate (BMR) and physical activity energy expenditure (PAEE), define exercise training zones, evaluate cardiovascular risk, and contribute to sleep-stage algorithms. Accordingly, even small inaccuracies may misinform users and practitioners, ultimately compromising the safety and efficacy of exercise programming, weight management strategies, and broader public health recommendations (Povea & Cabrera, 2018).

Although photoplethysmography (PPG) has become foundational in wearable cardiovascular monitoring, its accuracy is often undermined by biological variability. PPG works by emitting light (typically green or near-infrared) into the skin and detecting changes in reflectance or transmission caused by pulsatile blood volume fluctuations. This AC signal is superimposed on a larger, non-pulsatile DC component and is used to infer heart rate, arterial stiffness, or blood pressure. However, because the signal depends heavily on light–tissue interactions, it is highly susceptible to anatomical and physiological variation (Kim & Baek, 2023).

One key source of variation is skin tone, particularly melanin content. Melanin absorbs light in the visible spectrum, especially at green wavelengths, which are commonly used in commercial wearables. As demonstrated by Zonios et al. (2001) and Fallow et al. (2013), increased melanin reduces signal penetration into the blood-containing dermis, attenuating the AC component of the PPG signal. Ajmal et al. (2021) quantified this impact using Monte Carlo simulations, reporting up to a 15% drop in signal quality (AC/DC ratio) between Fitzpatrick type I and VI skin. Similarly, Boonya-Ananta et al. (2021) showed that with increasing melanin concentration, there was a consistent and measurable decline in PPG signal strength. Obesity and elevated body mass index (BMI) pose another significant challenge. Obesity alters several key dermal parameters: increased dermal thickness, greater trans-epidermal water loss (TEWL), reduced capillary density, and increased depth of arteries such as the radial artery. Ndumele et al. (2016) and Yosipovitch et al. (2007) have documented reduced dermal blood flow and altered skin hydration in obese individuals, both of which compromise the optical contrast needed for accurate PPG detection. Ajmal et al. (2021) observed over a 60% drop in AC/DC signal strength between normal-weight and morbidly obese individuals. Boonya-Ananta et al. (2021) confirmed these findings using hybrid FEM and Monte Carlo models, showing a 43% decrease in signal amplitude as BMI increased from 25 to 45 kg/m².

When skin tone and obesity co-occur, the combined effect is even more severe. Ajmal et al. (2021) noted a compounded degradation in signal quality when simulating a type VI skin tone with BMI >40, particularly in devices with shallow optical penetration depth like the Fitbit Versa 2. The AC/DC ratio dropped more than 60%, effectively limiting accurate cardiovascular inference for these populations. Additional demographic variables, such as age and sex, also influence PPG signal integrity. Age-related changes, such as decreased vascular compliance, increased arterial stiffness (Vlachopoulos et al., 2011), and thinning of the epidermis (Derraik et al., 2014), reduce the pulsatility and clarity of the PPG waveform. Sex differences in subcutaneous fat distribution may also influence the light path length and scattering behavior (Freedman et al., 2013). Moreover, epidemiological data further justify the importance of accurate, equitable PPG design.

Obesity is a known independent risk factor for multiple cardiovascular diseases (Ndumele et al., 2016; Wu et al., 2015), while underserved populations, who are more likely to present with darker skin tones or elevated BMI, are also at greater risk of cardiovascular morbidity. As wearable devices become central to health monitoring, failing to correct for these errors exacerbates existing healthcare disparities (Bent et al., 2020; Costin et al., 2007; Kollias & Baqer, 1985; Zonios et al., 2001). However, compared to skin tone, research on BMI's role in PPG error and its interaction is nascent and warrants deeper investigation.

In addition to participant-level sources of error, variation within individuals, particularly from fluctuations in heart rate and subtle motion artifacts, can further compromise PPG accuracy. Even when movement is minimized, as it is on a stationary cycle, exercise intensity alone can compromise PPG accuracy. As heart rate and vascular dynamics escalate during vigorous activity, the PPG signal becomes vulnerable to subtle motion artifacts and diminished signal fidelity (Fine et al., 2025). Studies using recumbent cycling have demonstrated that errors in HR measurement not only persist but actually increase with rising intensity, even with reduced limb movement (Hung et al., 2025). This phenomenon is thought to stem from the physical distortion of the sensor-skin interface and changes in blood volume waveform, both of which degrade optical sampling of pulse cycles (Maeda et al., 2011; Scardulla et al., 2023; Sirkiä et al., 2024). In some cases, wearables may even skip readings entirely when quality drops below internal thresholds, effectively masking rather than mitigating error (Park et al., 2025). Although our protocol controls for gross movement with a recumbent bicycle, intensity-related signal degradation remains a concern, underscoring the need to characterize error across the full range of exercise intensities.

Despite relying on the same underlying PPG technology, commercial wearables differ substantially in how they process optical signals, employing proprietary algorithms, sampling rates, sensor configurations, and noise-reduction strategies. These differences may result in device-specific error patterns, particularly when used across individuals with varying physiological and anatomical characteristics. Accordingly, the present study aimed to evaluate the accuracy of four leading wrist-worn HR monitors in a diverse sample, with particular attention to body mass index (BMI), skin pigmentation, and exercise intensity. We hypothesized that darker skin tones, quantified via the Fitzpatrick scale, would be associated with greater error in HR estimation, including systematic bias, absolute deviation, proportional error, and reduced reliability. Similarly, we expected higher BMI to yield elevated error across the same metrics, with the greatest inaccuracies occurring in individuals with both high BMI and darker skin tone.

Given the optical and physiological demands of PPG, we further anticipated that HR error would increase at higher exercise intensities. Finally, we hypothesized that error profiles would vary by exercise intensity and that the magnitude and direction of error would interact with participant characteristics such as body fat percentage (BF%), skin optical properties, age, gender, device location, and skinfold thickness, although these exploratory moderators were not central to our primary aims.

## Method

**Participants**

All study procedures received ethical approval from the Florida International University Institutional Review Board (IRB-22-0471-AM03; approval date: July 16, 2024) and adhered to the principles outlined in the Declaration of Helsinki (World Medical Association, 2013). Recruitment occurred through multiple channels, including flyers posted across campus, emails circulated through departmental listservs, and word-of-mouth referrals. Students enrolled in Kinesiology and Sport & Fitness courses were offered optional extra-credit for participation, with an equivalent non-research assignment available to those who opted out. Regardless of student status, each participant was compensated with a $50 Target gift card.

To ensure representation across key phenotypic characteristics relevant to wearable-sensor performance, a stratified sampling strategy was used based on BMI and skin tone. Eligibility criteria required individuals to self-identify as Hispanic adults between 18 and 50 years of age and to fall within Fitzpatrick skin types III-V (Fitzpatrick, 1988). Individuals were excluded if they did not meet these demographic criteria, reported any contraindications on the PAR-Q+ (Warburton et al., 2011), or had tattoos on the forearm that could compromise optical heart-rate sensing.

A power analysis conducted prior to data collection indicated that a minimum of 52 participants would be needed to detect a medium effect size ($f^2(v) = .20$) in a MANOVA with two groups and two dependent variables (absolute and directional error), assuming $\alpha = .05$ and power $(1–\beta) = .80$ (Bent et al., 2020). To mitigate potential data loss due to equipment malfunction or invalid sensor recordings, the recruitment target was increased to 58 individuals. During early data collection, the majority of volunteers had BMI values below 30. To maintain stratification goals, enrollment for this category was capped at 33 participants, after which only individuals with BMI $\geq$ 30 were enrolled until the final sample size was met.

The resulting sample consisted of 58 participants (31 women, 53.4%; 27 men, 46.6%), with a mean age of 23 years (SD = 5.91). Average BMI across the full sample was 29.73 (SD = 7.43). Participants with BMI < 30 (n = 33) had an average BMI of 24.24 (SD = 3.41), whereas those with BMI $\geq$ 30 (n = 25) had an average BMI of 36.97 (SD = 4.43). Skin tone distribution was as follows: Fitzpatrick type III (51.7%, n = 30), type IV (41.4%, n = 24), and type V (6.9%, n = 4).

**Design**

This study used a mixed factorial design. Participants were randomly assigned to wrist-placement configurations for the wearable devices and completed a single cycling session on a recumbent ergometer following a structured incremental-intensity protocol adapted from the ANSI/CTA-2065 (2018) consumer wearable testing standard.

**Measures**

Heart rate was recorded using a Polar H10 chest-strap electrocardiogram (ECG; Polar Electro, Kempele, Finland). The H10 has shown strong agreement with clinical ECG instrumentation in prior validation studies (Gilgen-Ammann et al., 2019). In addition to the criterion device, participants wore five wrist-mounted consumer wearables: the Apple Watch Series 8 (Apple Inc., Cupertino, CA), Garmin Forerunner 955 (Garmin Ltd., Olathe, KS), Fitbit Sense 2 (Google LLC, Mountain View, CA), Samsung Galaxy Watch 5 (Samsung Electronics, Seoul, South Korea), and the Empatica E4 (Empatica Inc., Cambridge, MA). The Empatica E4 was included during the early phase of recruitment but removed after the company discontinued support; therefore, its data were not retained for analysis. Wrist assignment for each device (left vs. right) was randomized across participants. Preliminary checks confirmed that wrist side did not influence outcomes ($ps > .05$).

Anthropometric characteristics were assessed using standard laboratory procedures. Height was measured with a wall-mounted stadiometer and weight with a calibrated digital scale. Body fat percentage (BF%) was estimated using skinfolds (Jackson & Pollock, 1978) and handheld bioelectrical impedance (Omron HBF-306C; Omron Healthcare, Kyoto, Japan). Wrist and forearm girths were obtained using a flexible anthropometric tape. Self-reported Fitzpatrick type was used in final analyses

**Procedure**

Testing sessions were approximately one hour in duration and were conducted in the kinesiology laboratory at Florida International University under controlled environmental conditions. Upon arrival, participants reviewed and signed informed consent documents and then completed demographic, medical history, and PAR-Q+ forms. Anthropometric measures (height, weight, BF%, and forearm/wrist dimensions) were obtained immediately afterward. Skin tone assessments, including Fitzpatrick classification and SFDS imaging, were completed next. Participants were then fitted with the Polar H10 chest strap and randomly assigned to a wearable-device placement configuration for the wrist-worn monitors. To minimize upper-body motion that could interfere with optical heart-rate sensing, participants were instructed to hold the side handles of the recumbent cycle ergometer throughout the protocol.

The exercise protocol consisted of a 10-minute cycling bout, shown in Figure 1. Participants first sat quietly for a 5-minute resting baseline. Subsequently, they completed alternating 2-minute cycling intervals at moderate and vigorous intensities. Target intensity zones were defined as 64–76% and 77–95% of age-predicted maximal heart rate, respectively, using the Tanaka equation (208 – 0.7 × age; Tanaka et al., 2001). Cadence was held at 85 revolutions per minute, and resistance was manually adjusted to maintain heart rate within the prescribed zone. The protocol concluded with a 5-minute seated recovery period. All procedures were preregistered on OSF: https://osf.io/9pzr2/overview?view_only=575bd3c163464781a0b600aee6a4d6c6.

**Figure 1**

*Cycling protocol used during the exercise trial.*

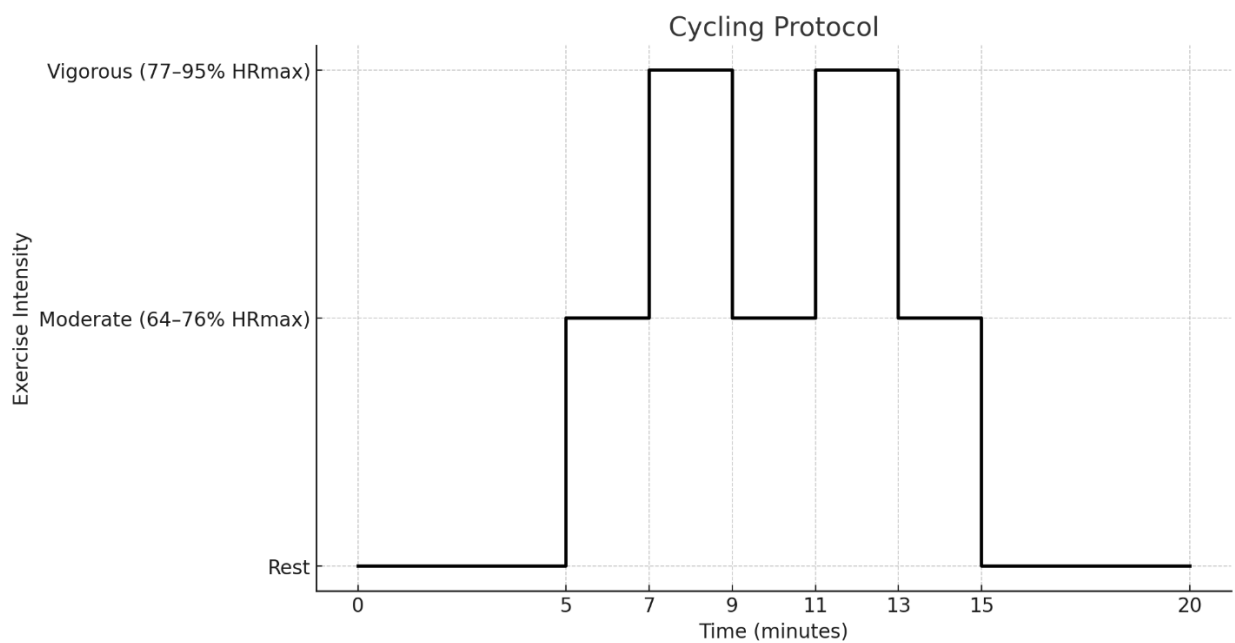

*Note*. After a 5-minute rest period, participants alternated between 2-minute intervals of moderate-intensity cycling (64–76% HRmax) and vigorous-intensity cycling (77–95% HRmax).

The sequence consisted of three moderate intervals and two vigorous intervals, followed by a 5-minute recovery period.

**Data Analysis**

All statistical analyses were conducted using R (v4.5.1) to evaluate the accuracy of wearable heart rate (HR) estimates across physiological and demographic variables. Four primary error metrics were derived for each device: Mean Absolute Error (MAE), Mean Absolute Percentage Error (MAPE), signed Bias, and Intraclass Correlation Coefficient (ICC) with the criterion HR (K5 metabolic system).

Preliminary inspection of metric distributions indicated skewness and bounded values, particularly for MAE, MAPE, and Bias, as well as potential heteroskedasticity across Fitzpatrick skin tone and BMI groups. Accordingly, non-parametric statistical methods were prioritized to ensure robustness and interpretability. Kruskal-Wallis tests were used to examine whether randomized device placement on the wrist (coded 1 through 6) significantly influenced error metrics within each device. In addition, to evaluate whether each device's error significantly deviated from perfect accuracy, Wilcoxon Signed-Rank Tests were conducted for MAE, MAPE, and Bias against a null value of 0, and for ICCs against a null value of 1.

Group-based comparisons of categorical variables (e.g., Fitzpatrick skin tone, BMI group, and gender) were analyzed using Kruskal-Wallis H tests and Mann-Whitney U tests. These rank-based tests were chosen for their independence from normality and homogeneity of variance assumptions. Eta-squared ($\eta^2$) was reported for Kruskal-Wallis tests, and rank-biserial correlation ($r_{rb}$) was used as an effect size for Mann-Whitney U tests. To test factorial interactions between Fitzpatrick skin tone and BMI group, a non-parametric Scheirer-Ray-Hare extension of the Kruskal-Wallis test was used for each metric. These models tested whether median errors differed across combinations of pigmentation and body composition without assuming parametric variance structures. Spearman rank-order correlations were used to examine associations between continuous predictors (e.g., BF% and age) and HR error metrics. This approach was selected to accommodate skewness and potential non-linear associations while preserving statistical power. Where significant associations emerged, $\rho^2$ was reported to indicate the proportion of variance accounted for by each predictor.

To assess the impact of continuous HR (exercise intensity) on HR error, linear regressions were conducted at the trial level, both pooled across devices and within device-specific subsets. Prior to modeling, absolute error (AE) and absolute percentage error (APE) were square root-transformed to reduce skew and stabilize variance. Model effect sizes were reported using standardized coefficients (r) and explained variance ($r^2$). Regressions were also repeated within each device to determine whether the influence of HR on error was device-specific.

To examine how heart rate error was influenced by individual physiological and phenotypic characteristics, a series of linear mixed-effects models were conducted separately for each device and each error metric (signed bias, $\sqrt{AE}$, $\sqrt{APE}$). These models used trial-level data (i.e., multiple HR estimates per participant per device), with participant ID included as a random intercept to account for repeated measures and inter-individual variability. Fixed effects included z-scored exercise intensity (K5 heart rate), z-scored BF%, and Fitzpatrick skin tone, along with all possible two- and three-way interaction terms. This hierarchical structure allowed the models to estimate systematic effects of physiological factors while accounting for participant-level differences in baseline accuracy and variance. Model performance was summarized using marginal $R^2$ (variance explained by fixed effects) and conditional $R^2$ (variance explained by both fixed and random effects). This approach enabled detection of complex interactive patterns (e.g., whether accuracy improved or worsened with intensity in individuals with high adiposity and darker pigmentation), while appropriately modeling the nested data structure. Device-level counts were reported after data cleaning. All statistical decisions were driven by the structure and distribution of the data. All data are provided at OSF:

https://osf.io/9pzr2/overview?view_only=575bd3c163464781a0b600aee6a4d6c6

## Results

### Overall Device Accuracy

To determine whether the wrist location of each device influenced performance, a series Kruskal-Wallis tests were conducted separately for each device (Apple, Garmin, Fitbit, Samsung) across the four error metrics (MAE, MAPE, bias, and ICC). The independent variable in each model was device position (coded 1 through 6), corresponding to randomized placement locations on the wrist. Across all tests, no significant main effects of placement location were observed (all $p$s > .50; full analysis available on OSF:

https://osf.io/9pzr2/overview?view_only=575bd3c163464781a0b600aee6a4d6c6).

To evaluate the overall accuracy of the four wearable devices, Wilcoxon Signed-Rank Tests were conducted comparing each device's MAE, MAPE, and bias against a theoretical value of zero (perfect accuracy), and ICCs against one (perfect agreement). Across all devices, results indicated systematic deviation from the criterion standard in at least three of the four metrics ($p$s < .001), with many tests yielding Wilcoxon statistics of zero, reflecting that all participants exhibited measurable error. Among devices, Apple and Garmin exhibited the lowest overall MAE and bias values. Additionally, all devices showed significantly reduced ICCs compared to perfect agreement, indicating some degree of reliability loss (see Table 1).

**Table 1**

*Overall device accuracy across four error metrics (MAE, MAPE, bias, and ICC) compared with criterion heart rate.*

| Device | Metric | Mean ± SD | Median (IQR) | Wilcoxon W (p) | N |
|---|---|---|---|---|---|
| Samsung | MAE | 3.38 ± 1.65 | 2.82 (1.62) | 0.0 (p < .001) | 33 |
| | MAPE | 3.10 ± 1.66 | 2.59 (1.69) | 0.0 (p < .001) | 33 |
| | Bias | 0.69 ± 1.69 | 1.17 (0.95) | 91.0 (p < .001) | 33 |
| | ICC | 0.96 ± 0.05 | 0.98 (0.04) | 0.0 (p < .001) | 33 |
| Apple | MAE | 2.51 ± 1.44 | 2.15 (1.63) | 0.0 (p < .001) | 35 |
| | MAPE | 2.36 ± 1.41 | 2.00 (1.45) | 0.0 (p < .001) | 35 |
| | Bias | 0.65 ± 1.13 | 0.59 (0.88) | 94.0 (p < .001) | 35 |
| | ICC | 0.98 ± 0.03 | 0.99 (0.02) | 0.0 (p < .001) | 35 |
| Garmin | MAE | 2.93 ± 1.39 | 2.52 (1.39) | 0.0 (p < .001) | 36 |
| | MAPE | 2.76 ± 1.42 | 2.51 (1.33) | 0.0 (p < .001) | 36 |
| | Bias | 0.83 ± 1.41 | 0.86 (0.94) | 82.0 (p < .001) | 36 |
| | ICC | 0.97 ± 0.03 | 0.98 (0.01) | 0.0 (p < .001) | 36 |
| Fitbit | MAE | 3.89 ± 2.28 | 3.16 (2.10) | 0.0 (p < .001) | 36 |
| | MAPE | 3.59 ± 2.31 | 2.60 (1.68) | 0.0 (p < .001) | 36 |
| | Bias | 0.59 ± 2.56 | 0.38 (1.66) | 259.0 (p = 0.252) | 36 |
| | ICC | 0.95 ± 0.07 | 0.98 (0.04) | 0.0 (p < .001) | 36 |

*Note.* MAE = mean absolute error; MAPE = mean absolute percentage error; ICC = intraclass correlation coefficient. Bias is defined as device heart rate minus criterion heart rate. Wilcoxon signed-rank tests (W) assessed differences from zero bias.

**Hypothesis Testing**

To examine whether the relationship between BMI category and device performance was moderated by Fitzpatrick skin tone, rank-based ANOVAs were conducted using a Scheirer-Ray-Hare approach for each error metric (MAE, MAPE, bias, and ICC). A significant interaction was observed for bias, $H(5, n = 140) = 4.80$, $p = .001$, and ICC, $H(5, n = 140) = 3.32$, $p = .007$. No significant interaction effects were found for MAE, $H(5, n = 140) = 1.85$, $p = .108$, or MAPE, $H(5, n = 140) = 2.05$, $p = .075$, although the latter approached significance. These findings suggest that combinations of high BMI and darker skin tone are uniquely associated with deviations in heart rate estimation, particularly in bias and reliability (See Table 2 AND Figure 2 for bias and ICC; all metrics are available on OSF: https://osf.io/9pzr2/overview?view_only=575bd3c163464781a0b600aee6a4d6c6). A follow-up series of nonparametric interaction-only Scheirer-Ray-Hare tests were conducted separately for each error metric specific to each device. Across all models, no significant interactions were observed between BMI group and Fitzpatrick skin tone (all $p$s > .37; detailed results are available on OSF: https://osf.io/9pzr2/overview?view_only=575bd3c163464781a0b600aee6a4d6c6). Given the low number of participants with Fitzpatrick skin tone 5, these results should be interpreted with caution, and additional examination of the main effects is warranted.

**Table 2**

*Interaction of BMI group and Fitzpatrick skin tone on bias and intraclass correlation coefficients (ICC).*

| BMI Group | Fitzpatrick | Mean | SD | Median | IQR | ICC Mean | ICC SD | ICC Median | ICC IQR | N |
|---|---|---|---|---|---|---|---|---|---|---|
| BMI <25 | 3 | -0.15 | 1.90 | 0.49 | 1.52 | 0.98 | 0.03 | 0.99 | 0.02 | 24 |
| BMI <25 | 4 | 0.08 | 1.65 | 0.48 | 0.95 | 0.98 | 0.05 | 0.99 | 0.01 | 24 |
| BMI <25 | 5 | -1.08 | 0.32 | -0.97 | 0.27 | 0.99 | 0.00 | 0.98 | 0.00 | 4 |
| BMI >30 | 3 | 1.11 | 1.67 | 0.84 | 1.5 | 0.95 | 0.06 | 0.98 | 0.06 | 50 |
| BMI >30 | 4 | 1.29 | 1.64 | 1.02 | 1.01 | 0.97 | 0.06 | 0.98 | 0.02 | 35 |
| BMI >30 | 5 | 0.71 | 0.44 | 0.58 | 0.43 | 0.97 | 0.01 | 0.97 | 0.01 | 3 |

*Note.* Values represent mean ± standard deviation (SD), median, and interquartile range (IQR). Nonparametric rank-based ANOVAs (Scheirer–Ray–Hare approach) were used to test for interactions. Bias values reflect directional error; ICC values indicate agreement with criterion HR.

**Figure 2**

*Bias and intraclass correlation coefficients (ICC) by BMI group and Fitzpatrick skin tone.*

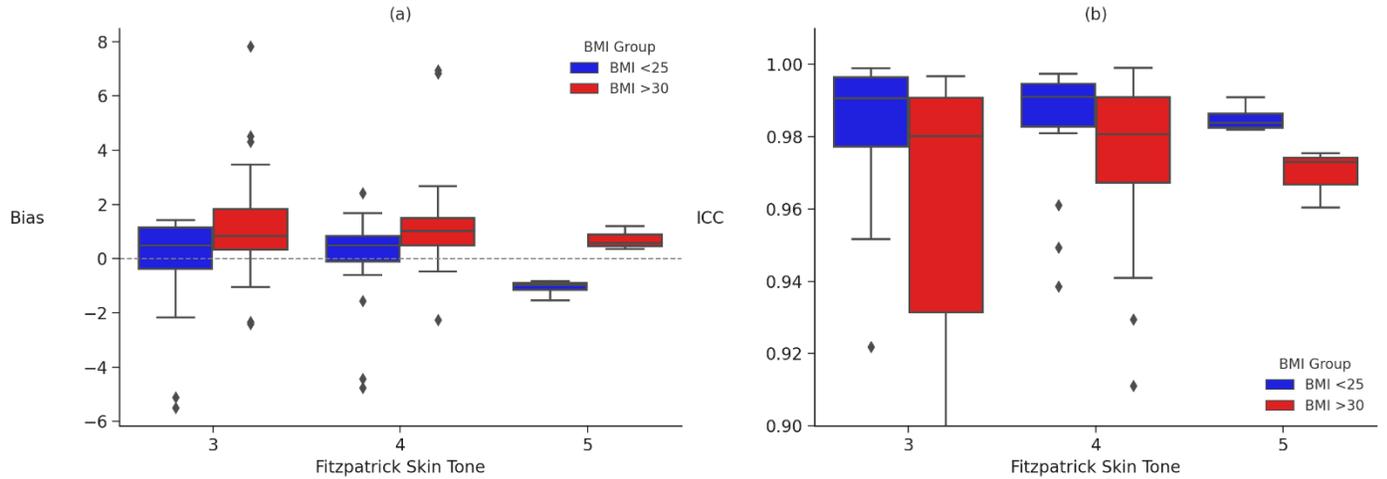

*Note.* Panel (a) shows directional bias values, and panel (b) shows ICC values across Fitzpatrick skin tones (III–V) for participants with BMI < 25 (blue) and BMI > 30 (red). Bias values reflect the direction of error (positive = overestimation, negative = underestimation). ICC values indicate agreement with criterion HR.

**Fitzpatrick**

To evaluate whether Fitzpatrick skin tone influenced device performance, Kruskal-Wallis tests were conducted for each error metric (MAE, MAPE, bias, and ICC). Significant main effects of skin tone were observed for MAPE, $H(2, n = 194) = 7.97, p = .019, \eta^2 = .215$, and bias, $H(2, n = 194) = 10.03, p = .007, \eta^2 = .263$, while MAE, $H(2, n = 194) = 5.89, p = .053, \eta^2 = .175$, and ICC, $H(2, n = 194) = 3.03, p = .220, \eta^2 = 0.102$, did not reach statistical significance. Post-hoc pairwise comparisons indicated that these effects were primarily driven by participants with Fitzpatrick skin type V, who exhibited significantly greater MAPE than those with skin type III, $U(n_1 = 11, n_2 = 102) = 292, p = .009, r_{rb} = .511$) and skin type IV, $U(n_1 = 11, n_2 = 81) = 205, p = .003, r_{rb} = .607$), as well as significantly lower bias than both Fitzpatrick III, $U(n_1 = 11, n_2 = 102) = 870, p = .003, r_{rb} = .517$, and Fitzpatrick IV, $U(n_1 = 11, n_2 = 81) = 684, p = .004, r_{rb} = .557$ (see Figure 3).

**Figure 3**

*Median and interquartile range of error metrics by Fitzpatrick skin tone.*

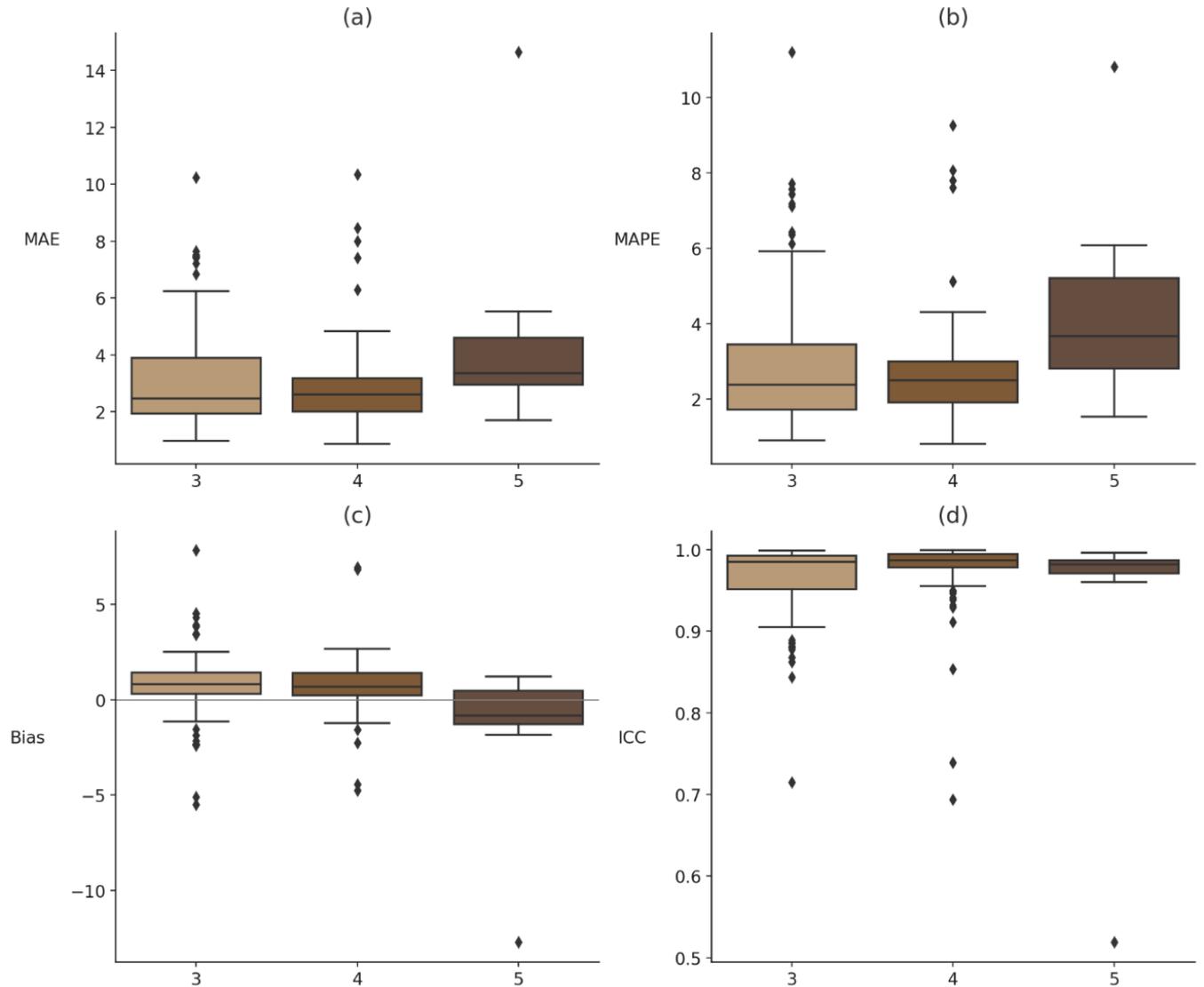

*Note.* Boxplots display the distribution of four error metrics, mean absolute error (MAE), mean absolute percentage error (MAPE), bias, and intraclass correlation coefficients (ICC), across Fitzpatrick skin tones III, IV, and V. Each box represents the interquartile range (IQR), the horizontal line indicates the median, and whiskers represent the range excluding outliers. Bias values reflect directional error, and ICC values reflect agreement with the criterion heart rate.

To evaluate whether Fitzpatrick skin tone influenced device performance at the individual device level, Kruskal-Wallis tests were conducted for each error metric (MAE, MAPE, bias, and ICC) within each device. A significant main effect of skin tone was observed for Samsung bias, $H(2, n = 47) = 6.64$, $p = .036$, $\eta^2 = .106$, while Apple bias trended toward significance, $H(2, n = 49) = 4.28$, $p = .118$, $\eta^2 = .078$. All other device-metric comparisons yielded non-significant results ($ps > .10$). Post hoc analyses indicated that the effect for Samsung bias was primarily driven by participants with Fitzpatrick skin type V, who showed significantly greater directional error than those with skin type III, $U(n_1 = 2, n_2 = 24) = 46$, $p = .025$, $r_{rb} = -.917$) and skin type IV, $U(n_1 = 2, n_2 = 21) = 41$, $p = .016$, $r_{rb} = -.952$). Critically, these biases are based on a very small sample size of participants with skin type V and should be interpreted with caution (see Figure 4; full device-level results are reported in OSF).

**Figure 4**

*Bland-Altman plots of device-level heart rate bias with 95% limits of agreement, stratified by Fitzpatrick skin tone.*

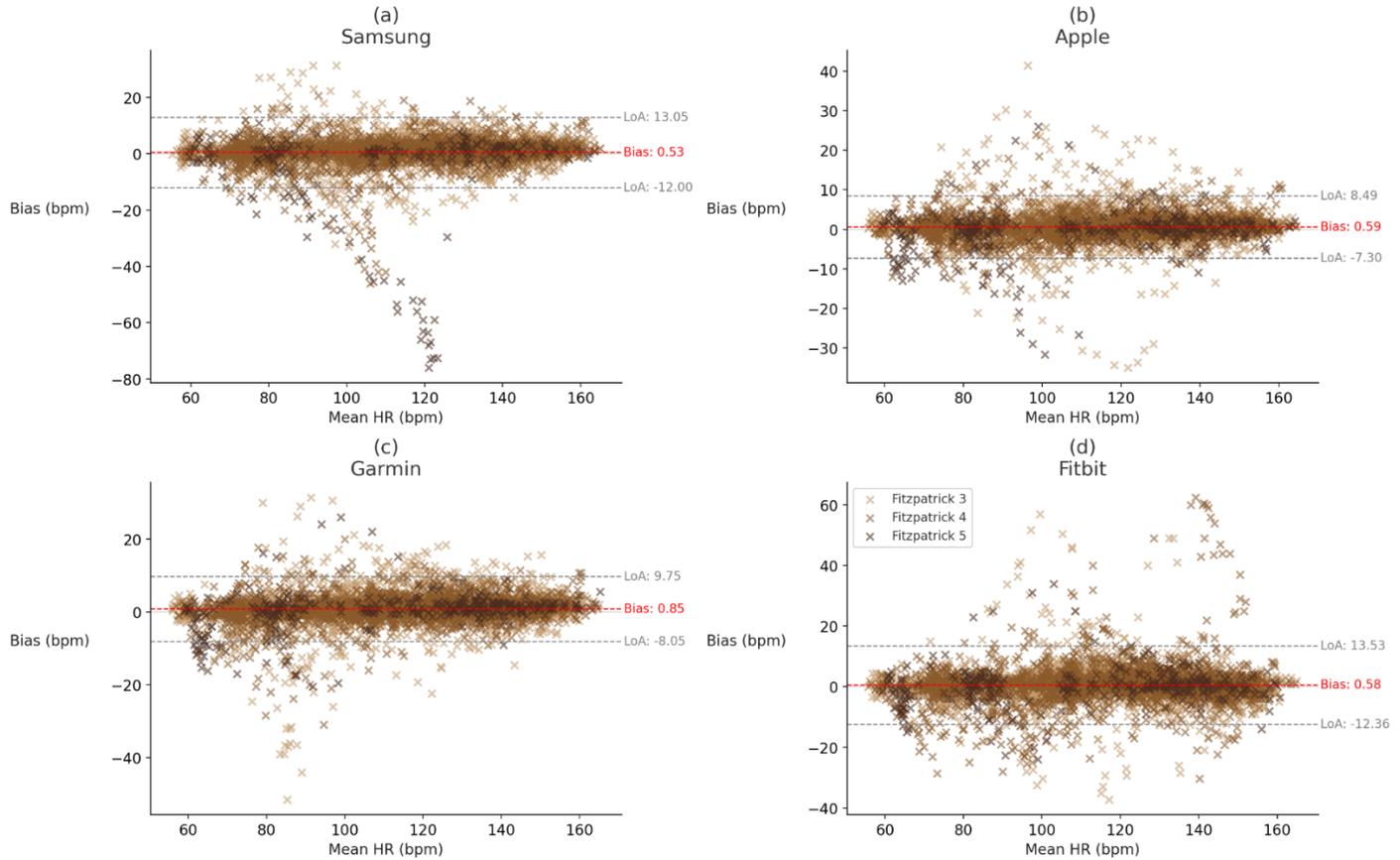

*Note.* Bland-Altman plots showing bias (in beats per minute) relative to criterion heart rate across (a) Samsung, (b) Apple, (c) Garmin, and (d) Fitbit devices. Data are stratified by Fitzpatrick skin tones III, IV, and V. Solid red lines indicate mean bias, and dashed gray lines indicate limits of agreement (LoA). Bias reflects directional error, with positive values indicating overestimation and negative values indicating underestimation.

**Body Composition**

To evaluate whether BMI category influenced device performance as originally hypothesized, Mann-Whitney U tests were conducted for each error metric (MAE, MAPE, bias, and ICC) comparing participants with BMI <25 and those with BMI >30. Significant effects of BMI were observed for bias, U($n_1$ = 52, $n_2$ = 88) = 1337, $p < .001$, $r_{rb}$ = .416), and ICC, U ($n_1$ = 52, $n_2$ = 88) = 3132, $p < .001$, $r_{rb}$ = –.369), indicating that participants with higher BMI exhibited greater directional bias and reduced measurement reliability. No significant differences were observed for MAE, $U(n_1 = 52, n_2 = 88) = 1986$, $p = .194$, $r_{rb} = .132$, or MAPE, $U(n_1 = 52, n_2 = 88) = 2253$, $p = .882$, $r_{rb} = .015$ (see Figure 5).

**Figure 5**

*Error metrics (MAE, MAPE, bias, ICC) by BMI group (< 25 vs. > 30) with 95% confidence intervals.*

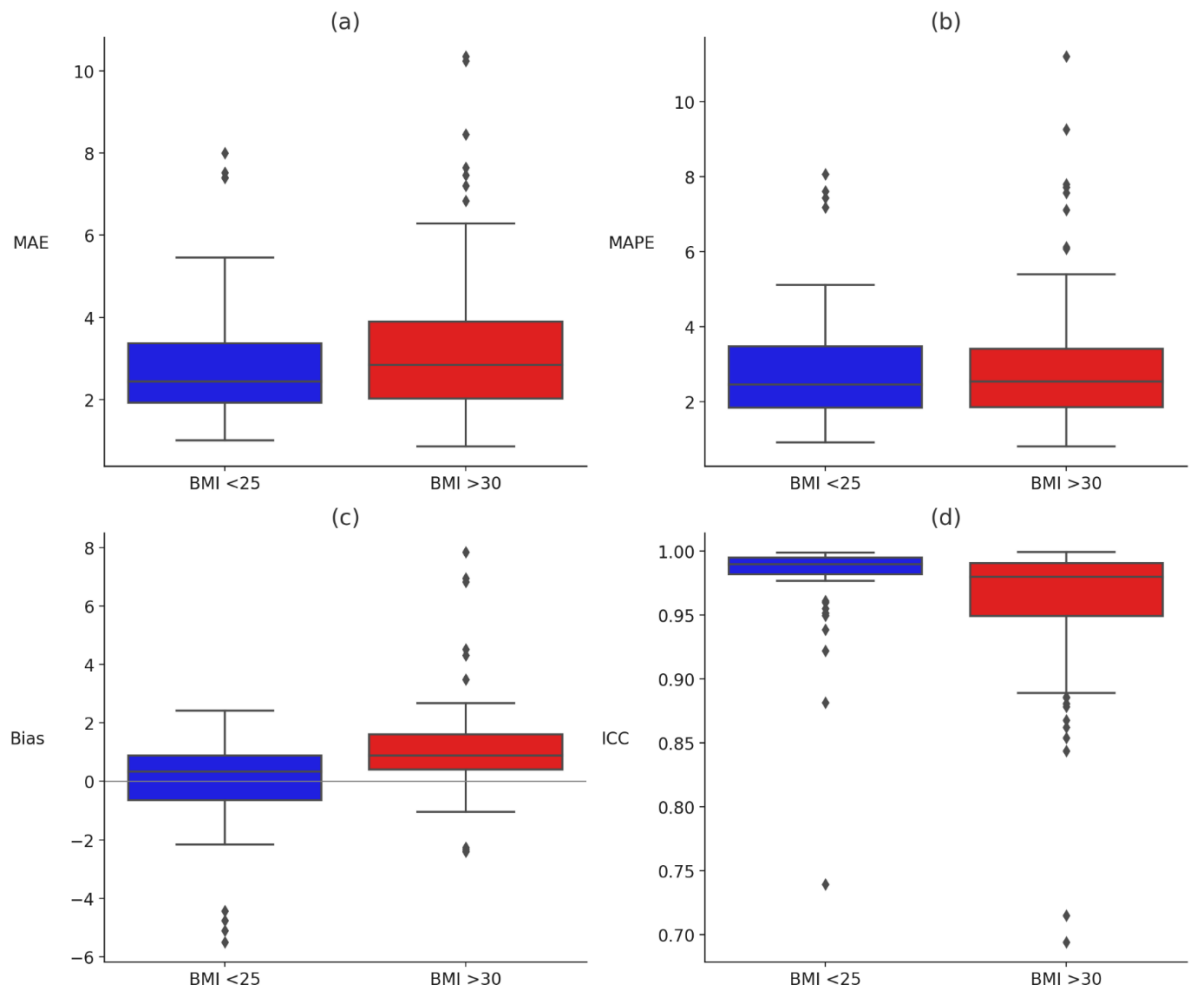

*Note.* Boxplots display the distribution of (a) mean absolute error (MAE), (b) mean absolute percentage error (MAPE), (c) bias, and (d) intraclass correlation coefficients (ICC) for participants with body mass index (BMI) < 25 and BMI > 30. Each box represents the

interquartile range (IQR), the horizontal line indicates the median, whiskers represent the range excluding outliers, and dots represent individual outliers. Error bars incorporate 95% confidence intervals. Bias reflects directional error, and ICC values indicate agreement with criterion heart rate.

To examine whether BMI category influenced device performance at the device level, Mann-Whitney U tests were conducted for each error metric (MAE, MAPE, bias, and ICC) within each device. Significant differences in bias were observed for the Apple, $U(n_1 = 22, n_2 = 13) = 72.0, p = .016, r_{rb} = .497$); Garmin, $U(n_1 = 23, n_2 = 13) = 63.0, p = .005, r_{rb} = .579$); and Fitbit, $U(n_1 = 23, n_2 = 13) = 81.0, p = .025, r_{rb} = .458$, devices, with participants with BMI > 30 exhibiting greater directional error in each case than participants with BMI < 25. Garmin also showed a significant difference in ICC, $U(n_1 = 23, n_2 = 13) = 212.0, p = .041, r_{rb} = -.418$, and the Apple, Fitbit, and Samsung devices trended toward significance for ICC, $U(n_1 = 22, n_2 = 13) = 200, p = .054, r_{rb} = -.399$; $U(n_1 = 23, n_2 = 13) = 208, p = .056, r_{rb} = -.391$; and $U(n_1 = 23, n_2 = 13) = 176.0, p = .094, r_{rb} = -.354$, respectively. All other comparisons were non-significant ($ps > .10$; see Figure 6; full device-level results are reported in OSF).

**Figure 6**

*Bland-Altman plots of heart rate bias for four wrist-worn devices compared to criterion measurement.*

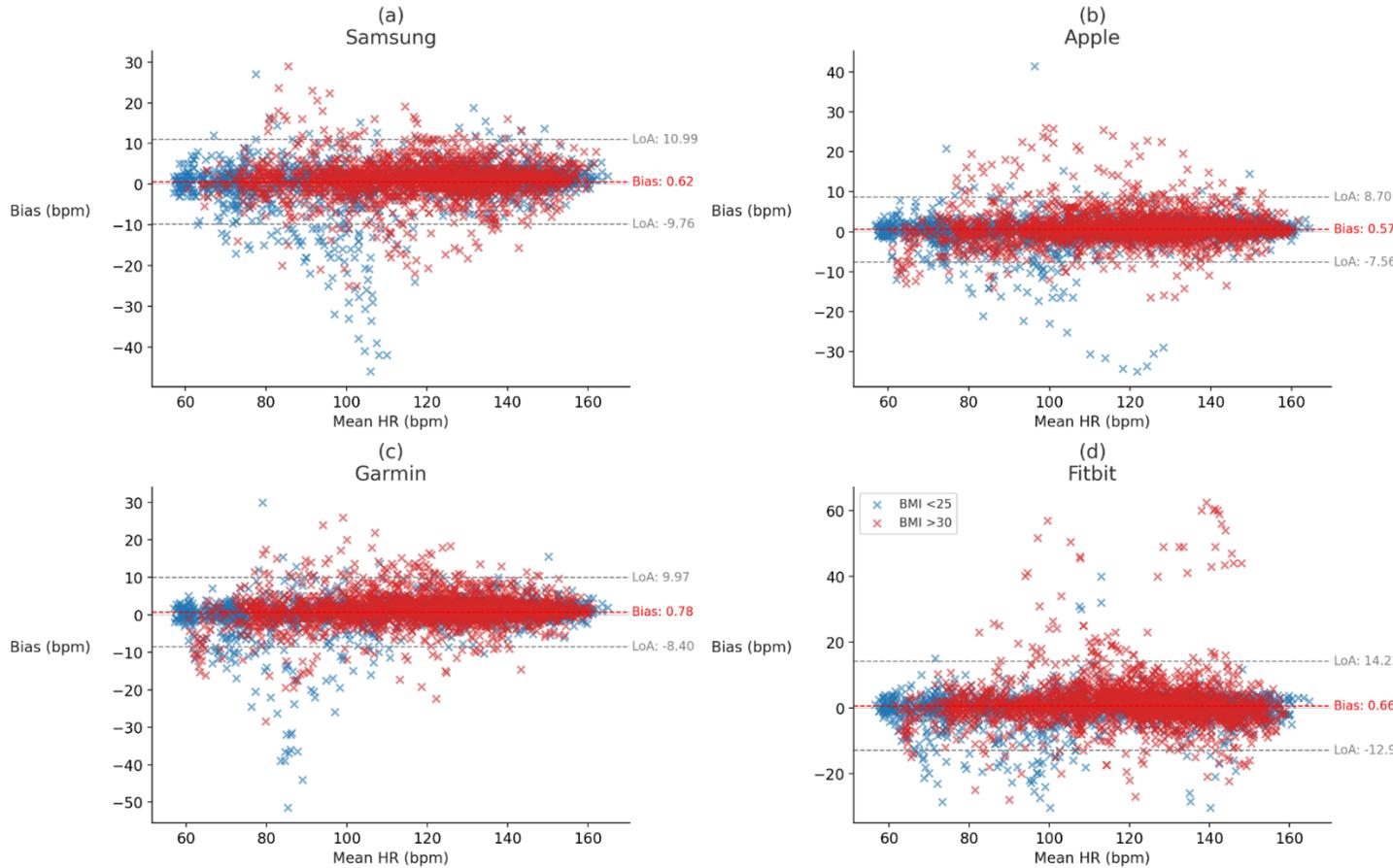

*Note.* Panels show bias in heart rate (device-criterion) for Samsung (a), Apple (b), Garmin (c), and Fitbit (d). Data points are color-coded by body mass index (BMI < 25 = blue; BMI ≥ 30 = red). The solid red line indicates the mean bias, and dashed lines indicate the 95% limits of agreement (LoA). Across devices, greater dispersion of error was observed among participants with higher BMI, with Fitbit showing the widest limits of agreement.

To evaluate BF% was associated with device performance, Spearman rank-order correlations were computed between BF% and each error metric (MAE, MAPE, bias, and ICC). Significant monotonic associations were observed for bias, $\rho$ (188) = .300, $p < .001$, $\rho^2 = 0.090$, and ICC, $\rho$ (188) = −.272, $p < .001$, $\rho^2 = .074$), indicating that higher BF% was associated with greater directional error and reduced reliability of heart rate measurement. No significant correlations were found between BF% and MAE, ρ (188) = 0.110, $p$ = .129, ρ² = .012 or MAPE, ρ (188) = .015, $p$ = .840, ρ² < 0.001; see Figure 7). Additional Spearman rank correlations were computed between BF% and each error metric (MAE, MAPE, bias, and ICC) for each device. Significant positive correlations were observed for bias in the Apple, ρ = .309, $p$ = .032); Garmin, ρ = .358, $p$ = .013; and Samsung, ρ = .316, p = .033, devices, indicating that participants with higher BF% exhibited greater directional error. A significant negative correlation was also observed between BF% and ICC for the Apple device, ρ = –.319, $p$ = .027), reflecting reduced agreement with the criterion standard as adiposity increased. All other device-metric correlations were non-significant (ps > .05; see Table 3; full device-level correlation results are available in OSF).

**Figure 7**

*Scatterplots of body fat percentage (BF%) and error metrics (MAE, MAPE, bias, ICC) with regression lines.*

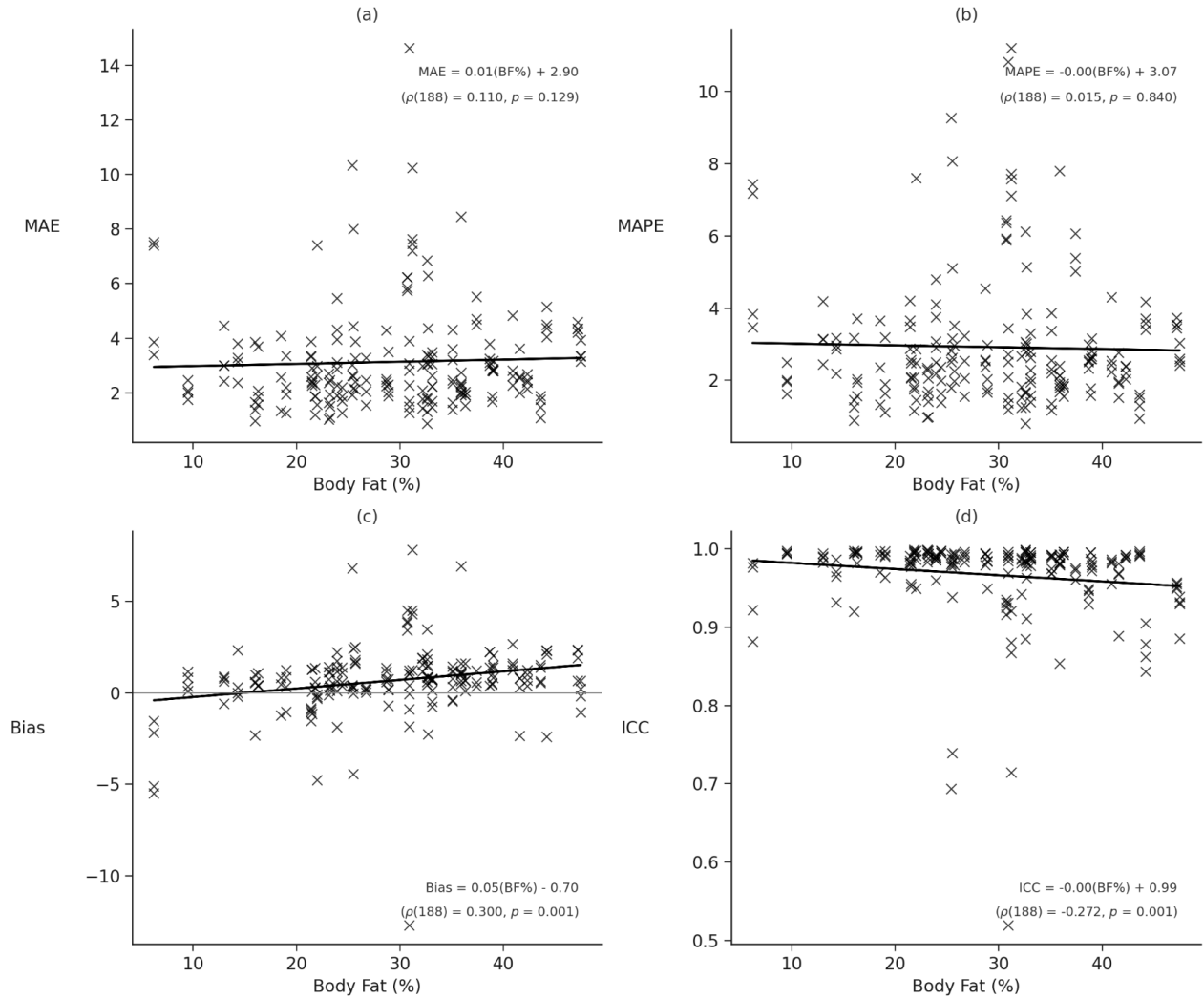

*Note.* Scatterplots depict relationships between BF% and error metrics across all devices. Regression lines are fit for each metric: (a) mean absolute error (MAE), (b) mean absolute percentage error (MAPE), (c) bias, and (d) intraclass correlation coefficient (ICC). Statistics shown within each panel are based on Spearman rank-order correlations with corresponding p-values.

**Table 3**

*Device-level Spearman correlations between body fat percentage (BF%) and error metrics (MAE, MAPE, bias, ICC).*

| Device | Metric | ρ | df | p-value | ρ² |
|---|---|---|---|---|---|
| Samsung | MAE | 0.073 | 44 | 0.631 | 0.005 |
| Samsung | MAPE | 0.008 | 44 | 0.9572 | 0.000 |
| Samsung | Bias | 0.316 | 44 | 0.0325 | 0.100 |
| Samsung | ICC | -0.257 | 44 | 0.0853 | 0.066 |
| Apple | MAE | 0.27 | 46 | 0.0638 | 0.073 |
| Apple | MAPE | 0.176 | 46 | 0.2324 | 0.031 |
| Apple | Bias | 0.424 | 46 | 0.0027 | 0.180 |
| Apple | ICC | -0.428 | 46 | 0.0024 | 0.183 |
| Garmin | MAE | 0.118 | 46 | 0.4239 | 0.014 |
| Garmin | MAPE | 0.002 | 46 | 0.9904 | 0.000 |
| Garmin | Bias | 0.508 | 46 | 0.0002 | 0.258 |
| Garmin | ICC | -0.262 | 46 | 0.0719 | 0.069 |
| Fitbit | MAE | -0.024 | 46 | 0.8726 | 0.001 |
| Fitbit | MAPE | -0.102 | 46 | 0.4912 | 0.010 |
| Fitbit | Bias | 0.114 | 46 | 0.4398 | 0.013 |
| Fitbit | ICC | -0.145 | 46 | 0.3247 | 0.021 |

*Note.* ρ = Spearman correlation coefficient; df = degrees of freedom. Positive correlations indicate that higher BF% was associated with increased error or bias; negative correlations indicate that higher BF% was associated with reduced agreement (ICC). Significant correlations (p < .05) are in bold.

**Exercise Intensity**

To examine whether heart rate (as measured by the K5 metabolic system) influenced wearable device performance, linear regressions were conducted using HR as a continuous predictor of three key error metrics: bias, square root-transformed Absolute Error ($\sqrt{AE}$), and square root-transformed Absolute Percentage Error ($\sqrt{APE}$). A statistically significant, albeit negligible, effect of HR on bias was observed, $b = -0.0044$, $t(20149) = -2.80$, $p = .005$, $r = -.020$, indicating a very slight tendency for directional error to decrease as heart rate increased. A small but significant negative association was also found between HR and $\sqrt{AE}$, $b = -0.0035$, $t(20149) = -15.0$, $p < .001$, $r = -.105$, suggesting that absolute error modestly decreased at higher intensities. A stronger relationship emerged between HR and $\sqrt{APE}$, $b = -0.0094$, $t(20149) = -$

43.6, $p < .001$, $r = -.283$, indicating that relative error significantly declined with increasing HR. Collectively, these findings suggest that wearable devices were more accurate during higher-intensity activity, particularly in terms of percentage error.

To further assess whether exercise intensity, as indexed by K5 heart rate (HR), influenced wearable device error, linear regressions were conducted for each device individually using HR as a continuous predictor of bias, square root-transformed Absolute Error ($\sqrt{AE}$), and square root-transformed Absolute Percentage Error ($\sqrt{APE}$). For the Samsung device, HR significantly predicted all three error metrics, indicating that directional, absolute, and relative errors decreased modestly with increasing HR. For the Apple device, HR did not significantly predict bias, but was significantly associated with $\sqrt{AE}$; no significant effect was observed for $\sqrt{APE}$. For Garmin device, HR was not a significant predictor of bias, but was significantly associated with both $\sqrt{AE}$ and $\sqrt{APE}$, suggesting modest improvements in accuracy with higher HR. For the Fitbit device, HR significantly predicted all three error metrics, with particularly strong reductions in relative error. Collectively, these findings suggest that HR-dependent reductions in error were device-specific, with Samsung showing consistent, though modest, HR-related improvements across all error metrics, while Apple showed more isolated effects, Garmin showed mixed effects, and Fitbit demonstrated the strongest HR-related improvement in relative accuracy. Full regression results are presented in Table 4 (see also Figure 6 for Bland-Altman plots).

**Table 4**
*Linear regression of heart rate (K5 criterion) predicting wearable device error metrics (bias, $\sqrt{AE}$, $\sqrt{APE}$).*

| Device | Metric | df | b | r² | r | p-value |
|---|---|---|---|---|---|---|
| Samsung | Bias | 4913 | -0.0121 | 0.0022 | -0.0471 | $p < .001$ |
| | √(AE) | 4913 | -0.0015 | 0.0019 | -0.0431 | $p = .002$ |
| | √(APE) | 4913 | -0.0080 | 0.0551 | -0.2347 | $p < .001$ |
| Apple | Bias | 5085 | -0.0003 | 0.0000 | -0.0017 | $p = .905$ |
| | √(AE) | 5085 | -0.0047 | 0.0270 | -0.1643 | $p < .001$ |
| | √(APE) | 5085 | -0.0098 | 0.1119 | -0.3344 | $p < .001$ |
| Garmin | Bias | 5076 | 0.0144 | 0.0062 | 0.0790 | $p < .001$ |
| | √(AE) | 5076 | -0.0045 | 0.0229 | -0.1514 | $p < .001$ |
| | √(APE) | 5076 | -0.0103 | 0.1104 | -0.3322 | $p < .001$ |
| Fitbit | Bias | 5069 | -0.0199 | 0.0056 | -0.0747 | $p < .001$ |
| | √(AE) | 5069 | -0.0032 | 0.0071 | -0.0841 | $p < .001$ |
| | √(APE) | 5069 | -0.0096 | 0.0661 | -0.2571 | $p < .001$ |

*Note.* df = degrees of freedom; b = unstandardized regression coefficient; r² = variance explained; r = correlation coefficient. √AE = square root-transformed absolute error; √APE = square root-transformed absolute percentage error.

**Additional Variables**

   **Age**

To examine whether age was associated with device performance, Spearman rank-order correlations were computed between age and each error metric (MAE, MAPE, bias, and ICC). A small but statistically significant correlation was found for ICC, $\rho(184) = -.207$, $p = .005$, $\rho^2 = .043$, indicating that older participants tended to have slightly lower agreement with the criterion measure. No significant associations were observed for MAE, $\rho(184) = -.097$, $p = .191$, $\rho^2 = 0.009$; MAPE, $\rho(184) = -.046$, $p = .528$, $\rho^2 = 0.002$; or bias, $\rho(184) = -.098$, $p = .188$, $\rho^2 = .010$. Device-level correlations showed no significant associations between age and any error metric (all *ps* > .05), suggesting that the observed reduction in ICC with age may reflect a small general trend rather than device-specific variation. Full correlation statistics by device are available in OSF.

   **Gender**

To examine whether gender was associated with wearable device accuracy, Mann–Whitney U tests were conducted comparing error metrics (MAE, MAPE, bias, and ICC) between male and female participants. A significant effect of gender was observed for ICC, $U(n_1 = 96, n_2 = 88) = 3104.5, p = .004, r_{rb} = –.230$), indicating that female participants had significantly lower HR reliability than male participants. No significant gender differences were found for MAE, $U = 3825.5, p = .437, r_{rb} = –.062$; MAPE, $U = 3942.5, p = .648, r_{rb} = –.035$), or bias, $U = 3675.5, p = .228, r_{rb} = –.092$). Device-level gender comparisons did not yield any statistically significant differences in error metrics ($ps > .05$), although several comparisons showed small-to-moderate effect sizes. Full Mann-Whitney U results by device are available in OSF.

**Hierarchical Linear Modeling**

To examine whether exercise intensity, body composition, skin tone, and their interactions impacted HR error in the four tested wearables, a series of hierarchical linear mixed-effects models were conducted. Models were nested within participant (random intercept) and included standardized K5 heart rate (z-scored), BF% (z-scored), Fitzpatrick skin tone, and all two- and three-way interactions as fixed effects. Each model targeted one of three dependent variables: signed residuals, square root-transformed absolute error ($\sqrt{AE}$), or square root-transformed absolute percentage error ($\sqrt{APE}$). Model fits indicated that marginal $R^2$ values ranged from 0.011 to 0.113, reflecting the proportion of variance explained by the fixed effects alone (see Table 5). Conditional $R^2$ values, which include both fixed and random effects, ranged from 0.07 to 0.26, suggesting substantial unaccounted for inter-individual variability and underscoring the importance of accounting for participant-level nesting in HR error analyses.

**Table 5**

*Marginal and conditional R² values for wearable device error metrics (bias, √AE, √APE).*

| Device | Metric | Marginal R² | Conditional R² |
|---|---|---|---|
| Samsung | Bias | 0.05 | 0.17 |
| | √(AE) | 0.01 | 0.17 |
| | √(APE) | 0.05 | 0.19 |
| Apple | Bias | 0.02 | 0.07 |
| | √(AE) | 0.04 | 0.20 |
| | √(APE) | 0.11 | 0.26 |
| Garmin | Bias | 0.05 | 0.09 |
| | √(AE) | 0.03 | 0.17 |
| | √(APE) | 0.11 | 0.24 |
| Fitbit | Bias | 0.02 | 0.12 |
| | √(AE) | 0.01 | 0.16 |
| | √(APE) | 0.07 | 0.21 |

*Note.* Marginal R² indicates variance explained by fixed effects; conditional R² indicates variance explained by both fixed and random effects. √AE = square root–transformed absolute error; √APE = square root–transformed absolute percentage error. AIC and BIC values are available in OSF.

For the Apple device, significant 3-way interactions between exercise intensity, BF%, and Fitzpatrick skin tone were observed for both bias, b = 0.27, t(4947) = 2.69, p = .007) and √(APE) (b = –0.04, t(4968) = –2.22, p = .026. There was also a significant two-way interaction between Fitzpatrick and exercise intensity for √(AE), b = –0.04, t(4963) = –2.96, p = .003. Taken together, these results indicate that Apple's HR error was moderated by both skin tone and body composition, with combined effects of higher exercise intensity, BF%, and pigmentation tending to increase directional bias, but reduce relative error (see COOL 3D model and supplemental materials for all model predictors in OSF).

For the Garmin device, all three metrics demonstrated significant three-way interactions between exercise intensity, BF%, and Fitzpatrick skin tone bias, b = 0.51, t(4935) = 4.46, p <

.001; √(AE), $b = –0.06$, $t(4962) = –3.07$, $p = .002$; and √(APE), $b = –0.07$, $t(4963) = –3.85$, $p < .001$. These results suggest that Garmin's HR error was highly sensitive to the combined influence of adiposity, pigmentation, and exercise intensity. Specifically, individuals with higher BF% and darker skin tones exhibited greater directional overestimation at higher intensities, but simultaneously experienced improvements in absolute and relative error, highlighting a nuanced interaction pattern where precision improved even as bias persisted (see COOL 3D model and supplemental materials for all model predictors in OSF).

For the Fitbit device, all three metrics bias, $b = 0.73$, $t(4957) = 5.16$, $p < .001$; √(AE), $b = –0.05$, $t(4950) = –2.61$, $p = .009$; √(APE): $b = –0.07$, $t(4950) = –3.51$, $p < .001$ indicated significant two-way interactions of exercise intensity and Fitzpatrick skin tone. Together, these results suggest that while Fitbit devices exhibited biased HR estimation as a function of skin tone and intensity, they also demonstrated improvements in absolute and relative accuracy under higher-intensity conditions for individuals with higher pigmentation (see COOL 3D model and supplemental materials for all model predictors in OSF).

For the Samsung device, all three metrics exhibited significant three-way interactions between exercise intensity, BF%, and Fitzpatrick skin tone bias, $b = –0.60$, $t(4799) = –3.40$, $p < .001$; √(AE), $b = 0.09$, $t(4795) = 3.66$, $p < .001$; and √(APE), $b = 0.07$, $t(4796) = 3.01$, $p = .003$. These findings suggest that Samsung devices tended to underestimate HR during high-intensity exercise in individuals with both higher adiposity and darker skin. However, both absolute and relative error increased in these same individuals, indicating that although the device's bias trended downward, its precision and percentage accuracy deteriorated under compounded physiological and pigmentation-related challenges (see COOL 3D model and supplemental materials for all model predictors in OSF).

## Discussion

The present study evaluated four wrist-worn PPG devices in a tightly controlled, low-motion cycling protocol and found that, although all devices deviated from the ECG criterion, absolute errors were small and agreement was generally moderate-to-high. From an applied perspective, these magnitudes are likely acceptable for most zone-based training and routine monitoring. Nonetheless, statistically reliable patterns tied to phenotype, particularly adiposity, and, to a lesser extent, skin pigmentation, were evident and align with the optical and physiological mechanisms described in prior work (Boonya-Ananta et al., 2021; Fine et al., 2021). When devices were pooled, darker skin tones were associated with greater error in the hypothesized direction (Fitzpatrick III-V; pooled test significant), consistent with optical absorption by melanin in the green band and reduced pulsatile content reaching the photodiode (Zonios et al., 2001; Fallow et al., 2013; Hung et al., 2025). In contrast, pooled comparisons by BMI category did not reach significance; however, several device-level contrasts indicated small directional biases with higher adiposity (e.g., Ajmal et al., 2021; Boonya-Ananta et al., 2021). Device-specific analyses yielded fewer skin-tone effects, and the cell size of the Fitzpatrick V subgroup was small, warranting cautious interpretation. Importantly, regression analyses indicated that the relative error decreased with increasing intensity across devices, suggesting that PPG systems may perform comparatively better when stroke volume and pulsatility increase (Maeda et al., 2011; Garet et al., 2005). This observation aligns with intensity-linked improvements in waveform salience, despite minimized motion in our protocol (Maeda et al., 2011; Fallow et al., 2013; Bent et al., 2020). These observations reinforce that commercial wearables work well on average under favorable conditions, while still exhibiting phenotypic sensitivity that can matter in specific contexts.

For most recreational and applied sport contexts, these devices likely deliver "good-enough" HR, especially when users make many physical activity decisions (e.g., zone checks, interval pacing, recovery; Shcherbina et al., 2017; Ludwig et al., 2018; Povea & Cabrera, 2018). Second, even small, systematic biases that cluster within phenotypic groups matter for equity-relevant downstream uses. A relatively small and hard-to-detect directional bias of 0.5-1.5 bpm or a few percentage points of proportional error may be inconsequential for interval pacing yet can propagate in energy-expenditure models and threshold detections that compound small errors over time (Shcherbina et al., 2017; Garet et al., 2005; Ludwig et al., 2018). Our mixed-effects models explained little variance via fixed effects, with far larger conditional $R^2$ driven by participant-level random effects, indicating substantial individual differences not captured by the measured covariates. This pattern suggests that person-specific factors, beyond BMI, BF%, and skin tone, account for much of the residual error. A minority of participants disproportionately occupied the error tails. In our sample, several of these cases had higher BMI and darker skin tones. However, given the modest subgroup sizes and collinearity between Fitzpatrick 5 and high BMI in these individuals, this observation is hypothesis-generating rather than confirmatory. This pattern dovetails with simulation work and reinforces the need for algorithms that adapt to person-specific optical contexts rather than assuming uniform tissue optics (Ajmal et al., 2021; Boonya-Ananta et al., 2021; Fine et al., 2021).

Our findings extend the literature in three ways. First, by focusing on a Hispanic cohort stratified on BMI and Fitzpatrick types III-V, we provide population-relevant estimates in a group at elevated cardiometabolic risk and historically under-represented in validation datasets. Second, by combining nonparametric group tests, device-level contrasts, and hierarchical models on trial-level data, we show that: adiposity increases bias and decreases reliability in 3/4 devices, pigmentation exerts a smaller but directionally consistent effect when power is pooled, and relative accuracy tends to improve with intensity. Third, by quantifying how little of the total variance is explained by observable fixed effects, we highlight that current commercial systems work well on average but remain vulnerable to errors in certain individuals, which are not random with respect to phenotypic features.

    Our protocol deliberately emphasized internal validity. We minimized gross motion with recumbent cycling at a standardized cadence and arm stabilization, synchronized all wearable streams to an ECG chest-strap criterion, and pre-specified nonparametric tests suited to bounded, skewed error distributions. We then complemented those tests with mixed-effects models that respected within-person nesting, allowing us to separate systematic influences from participant-level heterogeneity without overfitting device noise.

    These design choices also define the study's boundaries. A recumbent, low-motion context approximates a best-case optical environment; however, it may not benefit from accelerometry-informed algorithms. In free-living use, wrist rotation, variable contact pressure, and ambient light will introduce additional noise and could amplify both overall error and phenotype-linked disparities. The trial was short and intensity-segmented, so longer bouts may reveal thermal or vasomotor drift that our protocol could not capture. Skin-tone representation was limited (Fitzpatrick V under-powered; type VI absent), which may limit detectable pigmentation effects. Lastly, we did not manipulate firmware or algorithms, and versions may have evolved during data collection, constraining the specificity of device-level inferences.

    Translationally, for practitioners and end-users, these devices are suitable for most training decisions, especially when corroborated by perceived exertion and performance cues. However, caution is warranted when applying rigid thresholds (e.g., HR-defined clinical alerts)

to individuals with higher adiposity and darker pigmentation. For manufacturers and methodologists, equity-focused refinements are both feasible and urgent. This includes expanding training datasets to over-represent phenotypic edge cases, incorporating user-specific priors with brief guided calibrations to reduce device–person mismatch, and reporting subgroup performance transparently with distributional metrics to surface hard-case behavior.

Future work should directly test the hypothesized compounding of pigmentation and adiposity by oversampling participants with Fitzpatrick V-VI at higher BMI. Continued research should further quantify the relative contributions of BF%, body composition, skin tone, exercise intensity, and wrist geometry. Validation and quantification should extend to free-living settings with motion proxies and contact-pressure sensing. Studies should also evaluate downstream consequences for caloric expenditure, time-in-zone, training load, and weight-management algorithms. Finally, characterizing individual error signatures could enable brief calibration procedures that tailor devices to the user.

## Conclusion

Contemporary wrist-PPG devices are accurate enough for most everyday training decisions, but small, systematic biases linked to adiposity, and to a lesser degree skin pigmentation, persist, with disproportionate effects likely concentrated among higher-risk users. Precision without equity is insufficient; engineering and validation pipelines should explicitly target phenotypic hard cases so that wearable cardiovascular monitoring is broadly useful and accessible.